\documentclass[
superscriptaddress,
reprint,
 amsmath,amssymb,
 aps,
]{revtex4-2}

\usepackage{upgreek}
\usepackage{graphicx}
\graphicspath{{figures/}}
\usepackage{dcolumn}
\usepackage{bm}
\usepackage{hyperref}
\usepackage[mathlines]{lineno}
\usepackage{ulem}
\usepackage{multirow}
\usepackage{booktabs}
\usepackage{caption}
\usepackage{subcaption}
\usepackage{siunitx}
\usepackage{xcolor,colortbl}
\usepackage{array}
\usepackage{tabularx}
\usepackage{footmisc}
\usepackage{threeparttable} 

\begin{document}


\title{Lifetimes of excited states in P$^-$, As$^-$ and Sb$^-$}

\author{J. Karls}
\affiliation{%
 Department of Physics, University of Gothenburg, SE-412 96 Gothenburg, Sweden
}%
\author{M. Björkhage}
\affiliation{%
 Department of Physics, Stockholm University, AlbaNova, SE-106 91 Stockholm, Sweden
}%
\author{M. Blom}
\affiliation{%
 Department of Physics, Stockholm University, AlbaNova, SE-106 91 Stockholm, Sweden
}%
\author{N. D. Gibson}
\affiliation{%
 Department of Physics and Astronomy, Denison University, Granville, Ohio 43023, USA
}%
\author{O. Hemdal Lundgren}
\affiliation{%
 Department of Physics, University of Gothenburg, SE-412 96 Gothenburg, Sweden
}%
\author{M. Ji}
\affiliation{%
 Department of Physics, Stockholm University, AlbaNova, SE-106 91 Stockholm, Sweden
}%
\author{M. K. Kristiansson}
\affiliation{%
 Department of Physics, Stockholm University, AlbaNova, SE-106 91 Stockholm, Sweden
}%
\author{D. Leimbach}
\affiliation{%
 Department of Physics, University of Gothenburg, SE-412 96 Gothenburg, Sweden
}%
\author{J. E. Navarro Navarrete}
\affiliation{%
 Department of Physics, Stockholm University, AlbaNova, SE-106 91 Stockholm, Sweden
}%
\author{P. Reinhed}
\affiliation{%
 Department of Physics, Stockholm University, AlbaNova, SE-106 91 Stockholm, Sweden
}%
\author{A. Ringvall-Moberg}
\affiliation{%
 Department of Physics, University of Gothenburg, SE-412 96 Gothenburg, Sweden
}%
\author{S. Rosén}
\affiliation{%
 Department of Physics, Stockholm University, AlbaNova, SE-106 91 Stockholm, Sweden
}%
\author{H. T. Schmidt}
\affiliation{%
 Department of Physics, Stockholm University, AlbaNova, SE-106 91 Stockholm, Sweden
}%
\author{A. Simonsson}
\affiliation{%
 Department of Physics, Stockholm University, AlbaNova, SE-106 91 Stockholm, Sweden
}%
\author{D. Hanstorp}
\affiliation{%
 Department of Physics, University of Gothenburg, SE-412 96 Gothenburg, Sweden
}%

\date{\today}

\begin{abstract}
Radiative lifetimes of three elements of the nitrogen group have been experimentally investigated at the Double ElectroStatic Ion Ring Experiment (DESIREE) facility at Stockholm University. The experiments were performed through selective laser photodetachment of excited states of P$^-$, As$^-$ and Sb$^-$ ions stored in a cryogenic storage ring.
The experimental results were compared with theoretically predicted lifetimes, yielding a mixture of very good agreements in some cases and large discrepancies in others. 
These results are part of our efforts to map out the lifetimes of all excited states in negative ions. 
This data can be used to benchmark atomic theories, in particularly with respect to the degree of electron correlation that is incorporated in various theoretical models. 
\end{abstract}

\maketitle


\section{\label{sec:introduction}Introduction}

Negative ions are fragile systems, where the long-range Coulomb potential is not dominant as in the case of positive or neutral systems. 
Instead, the leading long-range force in negative ions is polarization, i.e. electron correlation. 
This means that negative ions cannot be described using the independent particle model, but more advanced models are needed instead \cite{Pegg2004StructureIons}. 

As a consequence of the short-range potential, the binding energy of negative ions is about an order of magnitude smaller than in neutral atoms, and there are several elements that do not form stable negative ions at all \cite{Andersen1999BindingIII}. 
Further, there are typically only a few excited states in negative ions which generally have the same parity as the ground state. 
Only five cases have been observed in negative ions so far where the excited states have opposite parity with respect to the ground states \cite{Tang_Th,Bil_Os, Walter_Ce2, Walter_La, Tang_U}, which is a requirement for an optically allowed transition. Therefore,  traditional methods of optical spectroscopy, which have yielded a wealth of data of the properties of atoms and molecules, cannot be applied to negative ions.

The main parameter that has been used to experimentally extract information about negative ions is the electron affinity (EA), the energy that is released by attaching an additional electron to a neutral system, which has been determined for a majority of the elements \cite{Ning2022}. 
Since this atomic quantity has been the main means to test atomic theories, as for instance has been shown in the recent study of At$^-$ \cite{Leimbach2020TheAstatine}, there is a continued effort and interest to perform high precision EA measurements. One  such recently performed study is the high resolution determination of the electron affinity of oxygen \cite{Kristiansson2022High-precisionOxygen}. 
However, comparing the experimental and theoretical value of a single quantity presents a risk that the factors that have been neglected in a model have opposing effects and cancel each other out, thereby giving a false impression of a good theoretical approximation. 
Consequently, it is clearly of interest to pursue additional ways to validate theoretical models. 
Here, lifetimes of excited states in negative ions offer an additional atomic property that can be studied and have shown to be very sensitive to electron correlation \cite{vonHahn2016TheCSR}. 

\begin{figure}
    \centering
    \includegraphics[width=0.45\textwidth]{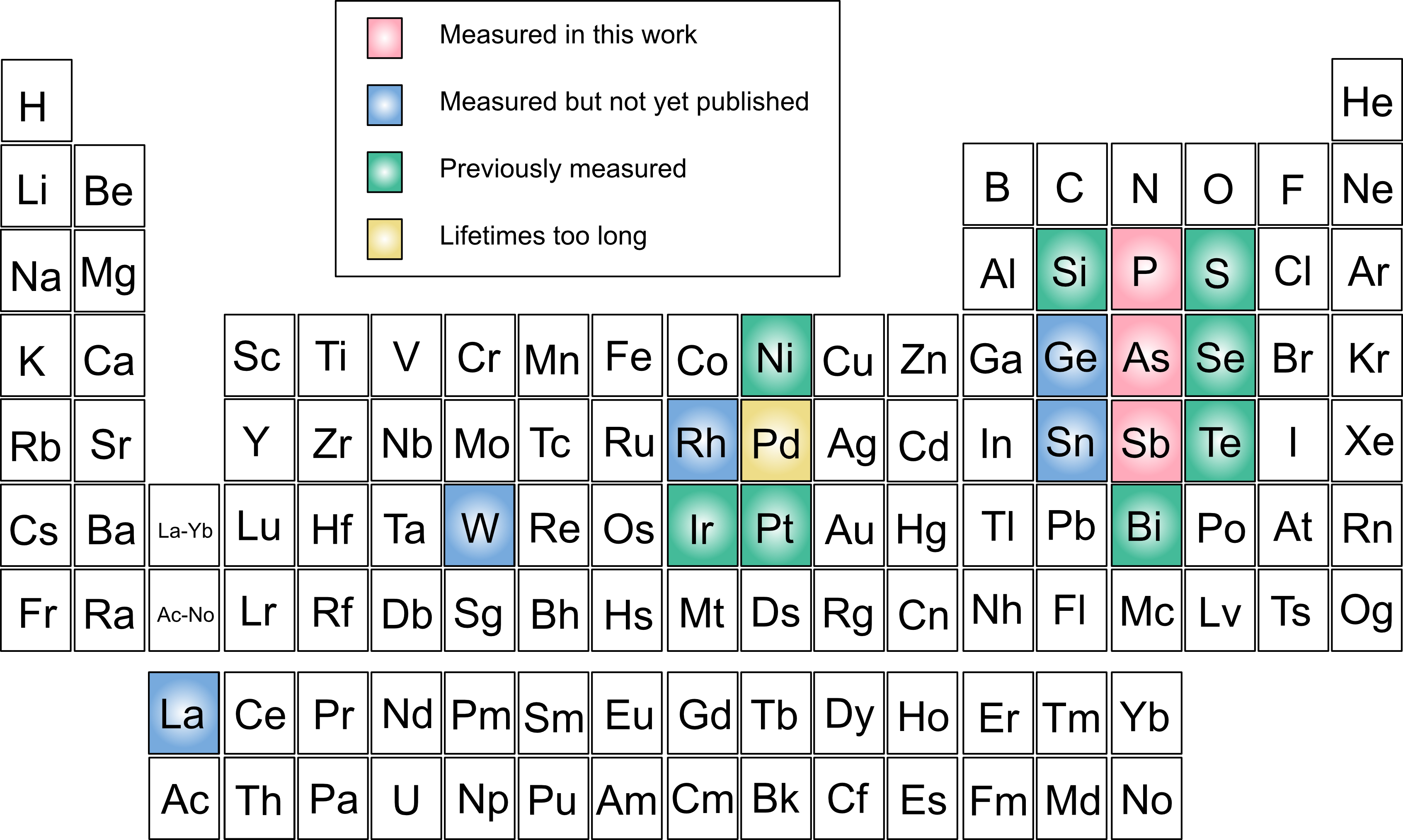}
    \caption{The periodic table. Elements where lifetimes of excited states were determined are indicated. Those presented in this work are marked in red, previously measurements are indicated in green \cite{Mull2021MetastableRing2,Kristiansson2022MeasurementBi-2,Kristiansson2021ExperimentalIr-2,Backstrom2015Storing-2,Andersson2006RadiativeIons,Ellmann2004RadiativeTe-2,Kaminska2016Lifetime,Chartkunchand2016Radiative}, elements where measurements have been performed but where the lifetimes were too long to be measured are in yellow and elements where measurements have been conducted but not yet published are blue.}.
    \label{fig:periodic_table}
\end{figure}

As most excited states in negative ions have the same parity as their corresponding ground states, their decay will be through M1 or E2 transitions, yielding very long lifetimes. 
This means that these transitions are not practically accessible in single-pass experiments, but the ions have to be stored for a longer period of time, which can be achieved at storage rings, such as the Cryogenic Storage Ring (CSR) \cite{vonHahn2016TheCSR}, or the Double ElectroStatic Ion Ring ExpEriment (DESIREE) \cite{Thomas2011TheStudies,Schmidt2013FirstDESIREE}. 
Lifetimes of excited states in anions of several elements have already been measured at DESIREE \cite{Kristiansson2021ExperimentalIr-2,Kristiansson2022MeasurementBi-2}, and at CSR \cite{Mull2021MetastableRing2}. 
Figure \ref{fig:periodic_table} shows an overview of the current status of the field. 

In this paper, we present measurement of lifetimes of excited states in P$^-$, As$^-$ and Sb$^-$ as part of our long term effort to map out the lifetimes of excited states in all elements that form negative ions. 
We hope that this can contribute to get a better understanding of the nature of negative ions and to the development of models that work for all negative systems describing both their structural and dynamical properties.




\section{\label{sec:experimentalmethods}Experimental methods}
Negative ions, originating from a cesium sputter ion source, were stored in the cryogenic storage ring DESIREE \cite{Thomas2011TheStudies,Schmidt2013FirstDESIREE}. 
In this ion source, negative ions are produced by sputtering with \SI{}{keV} energy Cs$^+$ ions, where the energy of the Cs$^+$ ions corresponds to a temperature of several million degrees. 
Hence, excited states in the negative ion will have high populations. 
After the production, the ions are accelerated and mass selected before being injected into the storage ring at a kinetic energy of \SI{20}{keV}.
Since the storage ring is held at 13 K, ions in the excited state will decay  until essentially the whole poplution is in the ground state.
The population of an excited state can then be probed by selective photodetachment as a function of the time after injection. 
In practice, the ions circulating in the storage ring are intersected by a laser beam detaching excited states in the ions, producing neutrals which are detected in a neutral particle detector \cite{Warbinek2019ADetector}.
The resulting photodetachment signal will yield an exponential decay over time due to the decreasing number of ions in the selected excited state, thereby allowing the lifetime of the excited state to be determined \cite{Backstrom2015Storing-_2}. 

Two different lasers have been used in the experiments: 
One \SI{10}{Hz} OPO with a tuning range of \SI{193}{} to \SI{2600}{nm}, and one \SI{1}{kHz} OPO with a tuning range of \SI{210}{} to \SI{2600}{nm}. 
The excited states were probed by using a laser with a photon energy smaller than the EA  but larger than the binding energy of the excited state of interest.
The decrease in total ion beam population has been monitored either by probing with a countinuous wave (cw) Ti:Sa laser with a photon energy larger than the EA, or by monitoring detachment through rest-gas collisions. 
Typically, ion currents of 1 nA or less were used to avoid ion-ion interactions in the ion beam. 

\begin{figure}
    \centering
    \includegraphics[trim={0cm 3cm 0cm 3.5cm},clip,width = 0.5\textwidth]{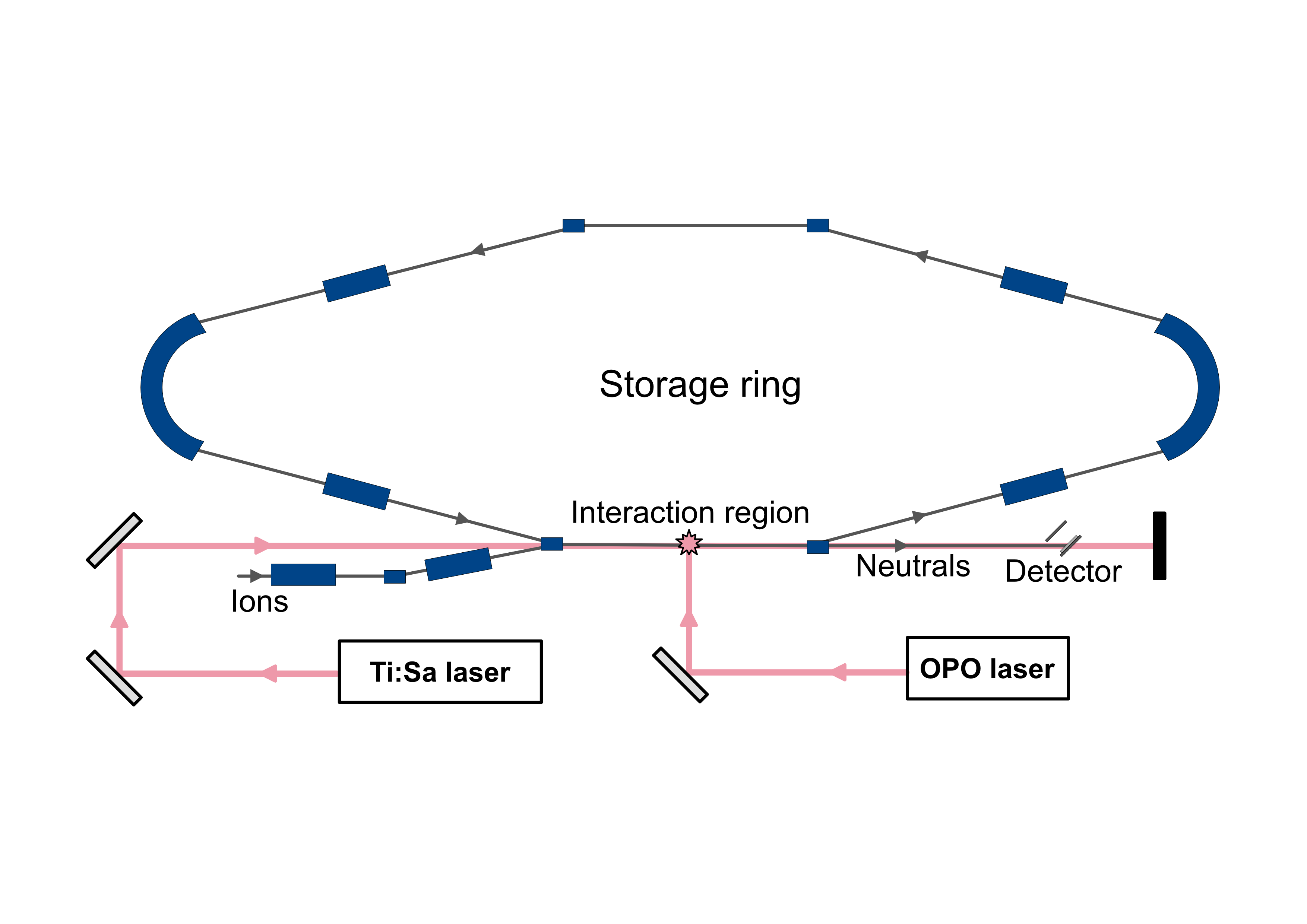}
    \caption{A schematic overview of the experimental setup. Ions are injected into the DESIREE storage ring where they circulate until they are neutralized either by a collision with the rest-gas or by a photodetachment process induced by light from one of the two lasers used in the experiments. 
     An Optical Parametric Oscillator (OPO) operating in a pulsed mode are intersected with the ion beam in a \SI{90}{\degree} angle, whereas a continuous wave Titanium:Sapphire (Ti:Sa) laser is collinear aligned with respect to the ion beam.}
    \label{fig:setup}
\end{figure}

\begin{figure*}
    \centering
    \begin{subfigure}[b]{0.3\textwidth}
        \centering
        \includegraphics[trim={1.5cm 0cm 2cm 0cm},clip,width=\textwidth]{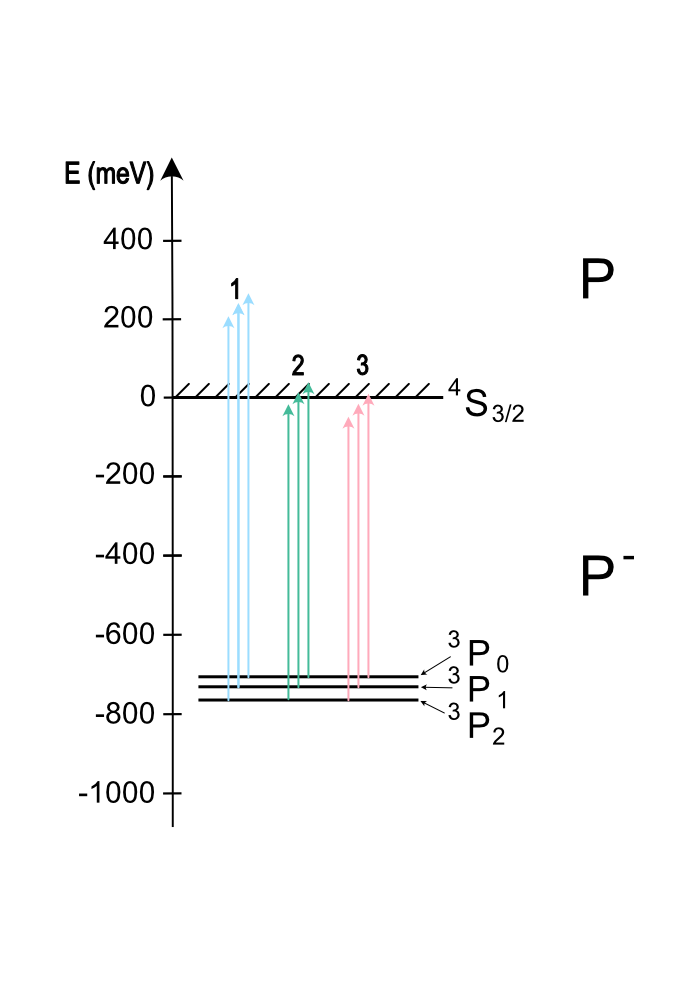}
        \caption{Schematic of P$^-$ energy levels.}
        \label{fig:P_levels}
    \end{subfigure}
    \hfill 
    \begin{subfigure}[b]{0.3\textwidth}
        \centering
        \includegraphics[trim={1.5cm 0cm 2cm 0cm},clip,width=\textwidth]{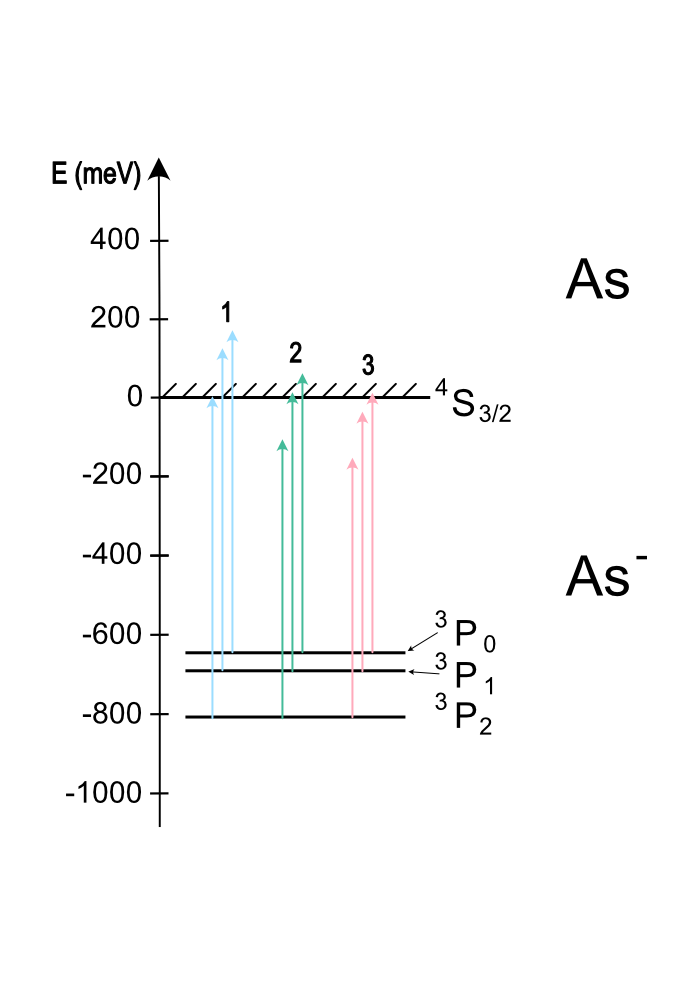}
        \caption{Schematic of As$^-$ energy levels.}
        \label{fig:As_levels}
    \end{subfigure}
    \hfill 
    \begin{subfigure}[b]{0.3\textwidth}
        \centering
        \includegraphics[trim={1.5cm 0cm 2cm 0cm},clip,width=\textwidth]{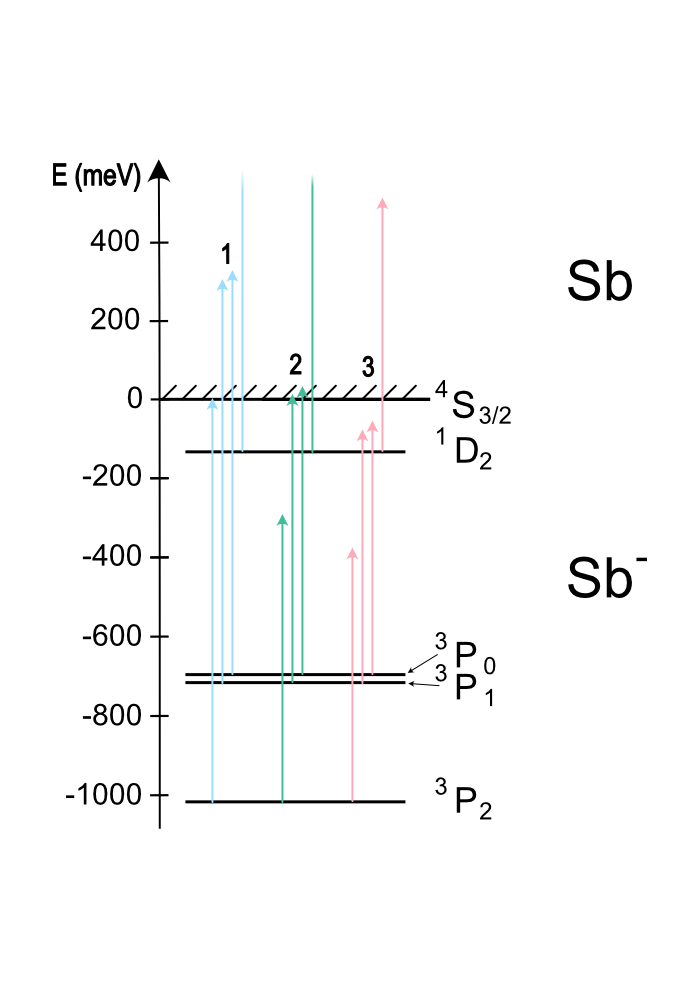}
        \caption{Schematic of Sb$^-$ energy levels.}
        \label{fig:Sb_levels}
    \end{subfigure}
    \caption{Schematics of energy levels for P$^-$, As$^-$, and Sb$^-$.}
    \label{fig:energy_levels}
\end{figure*}

\section{Results}
\begin{figure}[hb!]
    \centering
    \includegraphics[width=0.45\textwidth]{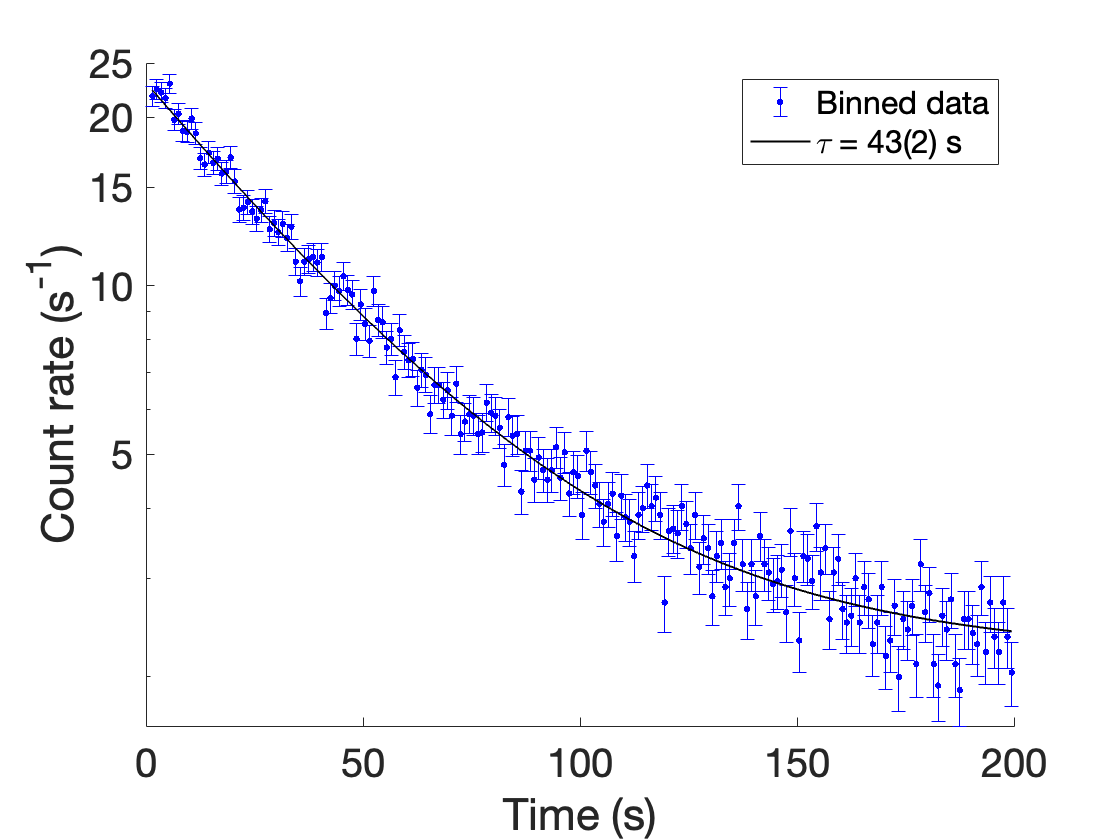}
    \caption{The decay curve of selective photodetachment of the $^3$P$_1$ state in As$^-$. A fit of the data yielded a lifetime of 43(2) s.}
    \label{fig:As_lifetime}
\end{figure}

Radiative lifetimes of the three nitrogen group elmements P$^-$, As$^-$ and Sb$^-$ have been determined. 
These elements all have a similar electronic structure, as shown in Fig. \ref{fig:energy_levels}. 
Here, blue arrows (1) indicate a photon energy used to photodetach all ions, irrespective of the state. 
The green arrows (2) indicate a photon energy that detaches all excited states but not the ground state. 
Finally, the red arrows indicate a photon energy that detaches only the least bound excited state.
A typical experiment would start with the latter, obtaining a single exponential curve, which allows to clearly identify the lifetime of the least bound excited state.
This lifetime can in turn be used for the determination of the lifetime of stronger bound excited states, where the measurement yields a double exponential decay due to the detachment of multiple excited states.
As an example, Figure \ref{fig:As_lifetime} shows the decay curve of selective photodetachment of the excited $^3$P$_1$ state in As$^-$. 
From the exponential decay fit, a lifetime of \SI{43(2)}{s} was extracted.

\begin{figure*}[]
    \centering
    \includegraphics[trim={3cm 0cm 3cm 0cm},clip,width=\textwidth]{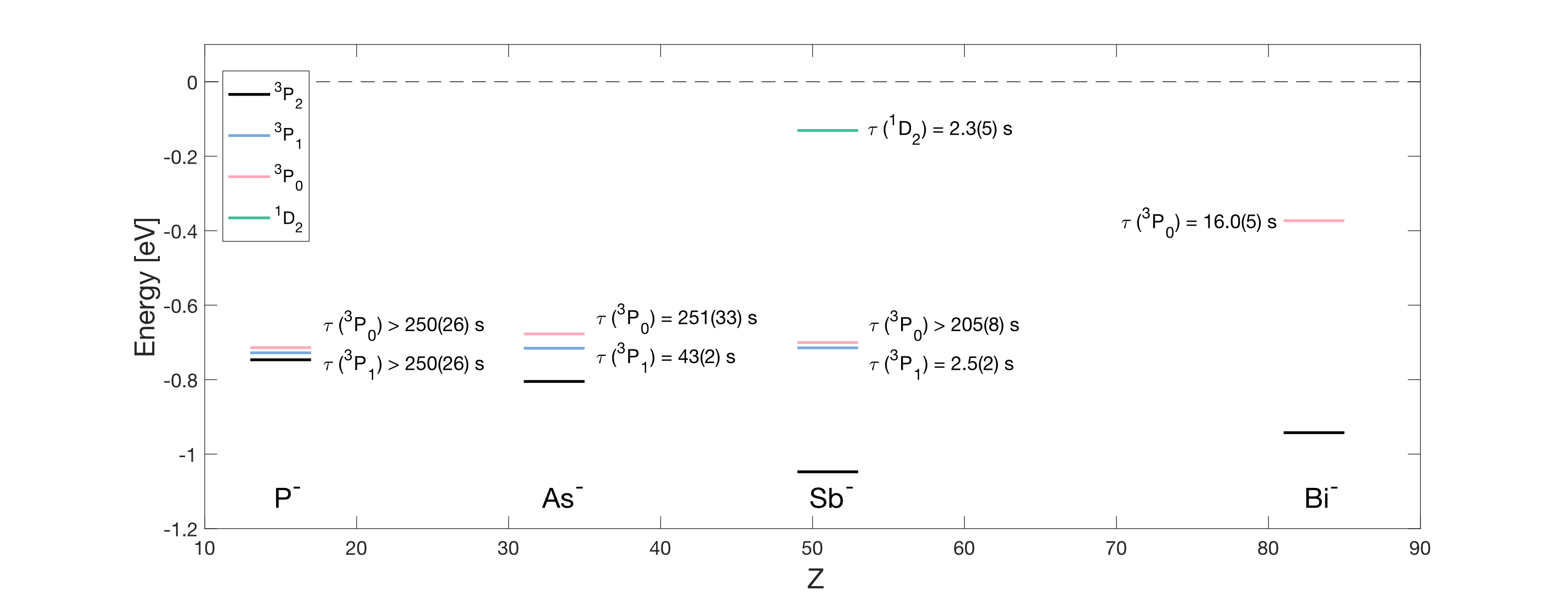}
    \caption{Lifetimes of excited states, presented in their corresponding energy level diagrams, for the non-radioactive elements in the nitrogen group that form stable negative ions. The lifetime of the $^3$P$_0$ in Bi$^-$ was measured previously by Kristiansson \textit{et al.} \cite{Kristiansson2022MeasurementBi-2}.}
    \label{fig:all_lifetimes}
\end{figure*}

The lightest of the studied elements, P$^-$, has two bound excited states, which both have lifetimes that exceed the storage time of negative ions at DESIREE. 
Hence, for the lifetimes of these two energy levels, a lower limit of \SI{250(26)}{s} was found. 
For As$^-$ we were able to measure the lifetimes of both of the excited states that exist in this ion, resulting in a lifetime of \SI{43(2)}{s} for the $^3$P$_1$ state and \SI{261(33)}{s} for $^3$P$_0$ state using the experimental procedure described above. 
Finally, for Sb$^-$ we were able to measured the lifetimes of the $^1$D$_2$ and $^3$P$_1$ states while for the lifetime of the $^3$P$_0$ it was only possible to determine a lower limit of the lifetime.
A summary of the results is presented in Figure \ref{fig:all_lifetimes}, and in  table \ref{tab:Lifetimes}, where they are also compared to lifetimes calculated by Su \textit{et al.} \cite{Su2019TheN-}.

\begin{table}[hb!]
    \centering
    \caption{Experimental and theoretical (\cite{Su2019TheN-}  lifetimes in  P$^-$, As$^-$ and Sb$^-$.}
    \begin{threeparttable}
         \begin{tabular}{@{}llll@{}}\toprule Element \hspace{0.2cm}& Energy level \hspace{0.2cm} &  $\tau_{\mathrm{Exp}}$ (s)\hspace{0.2cm}& $\tau_{\mathrm{Th}}$ (s) \cite{Su2019TheN-} \\\midrule 
        \multirow{3}{*}{P$^-$} & $^3$P$_2$ & - & - \\ 
         & $^3$P$_1$  & $> 250(26)$ & $7.457 \cdot 10^3$ \\     
         & $^3$P$_0$  & $> 250(26)$ & $3.291 \cdot 10^4$ \\  \midrule 
        \multirow{3}{*}{As$^-$} & $^3$P$_2$ & - & - \\ 
         & $^3$P$_1$  & 43(2) & $41.26$ \\  
         & $^3$P$_0$  & 251(33) & 722.3 \\  \midrule 
        \multirow{4}{*}{Sb$^-$} & $^3$P$_2$ & - & - \\ 
         & $^3$P$_1$  & 2.5(2) & 2.426 \\  
         & $^3$P$_0$  & $> 205(8)$  & $1.261 \cdot 10^4$\\  
         & $^1$D$_2$  & 2.3(5) & 1.188 \\ 
        \bottomrule 
        \end{tabular} 
    \end{threeparttable} 
    \label{tab:Lifetimes}
\end{table}

\section{Discussion}

In this work, we have measured lifetimes of the excited states of three elements in the nitrogen group, namely P$^-$, As$^-$ and Sb$^-$. 
While the nitrogen atom does not form a negative ion \cite{Andersen1999BindingIII, Cowan_1997}, the Bismuth anion was recently investigated by Kristiansson \textit{et al.} \cite{Kristiansson2022MeasurementBi-2}.

They obtained a lifetime of \SI{16.0(5)}{s} for the $^3P_0$ state, which is in good agreement with the theoretical value of Walter \textit{et al.} \cite{Walter2021} of \SI{16.5(7)}{s} as well as the value of Su \textit{et al.} \cite{Su2019TheN-} of \SI{15.2}{s}.
The last element in this group, moscovium, is purely radiogenic and has a lifetime of only  1.5 minutes. 
It belongs to the transuranium elements and no negative ions of this group of elements have so far been observed. 
With our results, we therefore complete the experimental investigation of the excited states of stable elements in the nitrogen group, as shown in Fig.\ref{fig:periodic_table},

In the following, we will compare our results with the theoretical lifetimes calulated by Su \textit{et al.} \cite{Su2019TheN-}, as shown in Table \ref{tab:Lifetimes}. 
In the case of the phosphorus anion, we were only able to determine a lower limit of \SI{250(26)}{s} for both of the two excited states, whereas the theoretical prediction is orders of magnitude larger.
The same is the case for the $^3P_0$ state in antimony, where a lower limit of \SI{205(8)}{s} was determined.
Hence, no conclusion of the accuracy of the calculated values can be drawn in these cases.
While the respective $^3P_1$ states of the arsenic and antimony anions agree very well with the theoretical prediction, for two cases there is a substantial discrepancy between theory and experiment: 
The $^3$P$_0$ state in As$^-$ shows an experimental lifetimes of just over 1/3 of the theoretical value, and for the $^1$D$_2$ state in Sb$^-$, the experimental value is twice the theoretical lifetime.
Figure \ref{fig:all_lifetimes} shows the lifetimes of the excited states, presented in their corresponding energy level diagrams, for the non radioactive elements in the nitrogen group that form stable negative ions.
An observation here is that there appears to be a good agreement between theory and experiment for the state with the lowest excitation, and that the larger discrepancy occurs for states with higher excitation energies. 
However, there are not sufficient data points for judging whether this is a trend or just a coincidence.



\section{Conclusions}

When comparing our result of P$^-$, As$^-$ and S$^-$, and  the result of Bi$^-$ of Kristiansson  \textit{et al.} \cite{Kristiansson2022MeasurementBi-2}, with the theoretical values from Su \textit{et al.} \cite{Su2019TheN-} and Walter \textit{et al.} \cite{Walter2021} we observed a mixed agreement.  
The good agreement for the states of lowest excitation points toward a solid theoretical  model, whereas the calculations for the P$_0$ state in As$^-$ and the $^1$D$_2$ state in Sb$^-$, need to be improved. 
Clearly, different states have different sensitivity to electron correlation effects. 
In order to advance the understanding of electron correlation, theory needs to include higher order excitations in the model, and corresponding studies for other groups in the periodic table have to be performed, which can be used as a tool to investigate shortcomings of theoretical models. We therefore plan to continue our work.


\begin{acknowledgments}
DH and HS acknowledge support from the Swedish Research Council under contracts 2020-03505 and 2022-02822. Financial support from the Swedish Research Council No. 2020-03505 is acknowledged. We want to thank the staff at DESIREE for their support during the beamtime.
\end{acknowledgments}

\bibliographystyle{unsrt}
\bibliography{references,references2}

\end{document}